\documentclass[12pt,preprint]{aastex}
\usepackage{natbib}
\usepackage{epsfig}
\usepackage{graphicx}
\input epsf.sty

\shorttitle{Long GRB durations}
\shortauthors{Janiuk \& Proga}

\begin{document}

\title{Low angular momentum accretion in the collapsar: how long
  can be a long GRB? }

\author{A. Janiuk\altaffilmark{1,2}, D. Proga\altaffilmark{1}}
\altaffiltext{1} {University of Nevada, Las Vegas, 4505 Maryland Pkwy,
  NV ~89154, USA}
\altaffiltext{2} {Copernicus Astronomical Center,
Bartycka 18, 00-716 warsaw, Poland}
\begin{abstract}
\end{abstract}

The collapsar model is the most promising scenario to explain the
huge release of energy associated with long duration gamma-ray-bursts
(GRBs). Within this scenario GRBs are believed to be powered by accretion
through a rotationally support torus or by fast rotation of a compact object. 
In both cases then, rotation of the progenitor star is one of the key
properties because it must be high enough 
for the torus to form, the compact object to rotate very fast, or both.
Here, we check what rotational properties a progenitor star must have in order
to sustain torus accretion over relatively long activity periods as observed 
in most GRBs. We show that simple, often cited, estimates of
the total mass available for torus formation and consequently
the duration of a GRB are only upper limits. We revise these estimates
by taking into account the long term effect that as the compact object 
accretes the minimum specific angular momentum needed for torus formation
increases. This in turn leads to a smaller
fraction of the stellar envelope that can form a torus.
We demostrate that this effect can lead to a significant,
an order of magnidute, reduction of the total energy
and overall duration of a GRB event. This of course can be
mitigated by assuming that the progenitor star rotates faster then
we assumed. However, our assumed rotation is already high compared
to observational and theoretical constraints.
 We estimate the GRB duration times, first by assuming 
a constant accretion rate, as well as by explicitly calculating the 
free fall time of the gas during the collapse.
We discuss the implications of our results.


\section{Introduction}

The collapsar model
for a gamma ray burst invokes a presence of an
accretion torus around a newly born black hole (Woosley 1995;
Paczy\'nski 1998;  MacFadyen \& Woosley 1999).
The accretion energy is being transferred to the jet that propagates through
the collapsar envelope and at some distance from the central engine is
responsible for producing gamma rays. This type of model is commonly
accepted as a mechanism for a long gamma ray burst production, because
the whole event can last as long as the fallback material from the
collapsar envelope is available to fuel the accretion disk or torus.

However, one should bear in mind that the rotating torus 
may form only when the substantial amount of specific angular momentum is
carried in the material 
(see e.g. Lee \& Ramirez-Ruiz 2006 for a recent study of this problem). 
This can be parameterized by the so called
critical specific angular momentum value, which is dependent on the
mass of the black hole, i.e. $l_{\rm crit}=2 R_{\rm g} c$,
where $R_{\rm g}$ is the gravitational radius. 
Because the black hole mass is not constant
during the collapsar evolution, but increases as the envelope 
material accretes onto it, the critical angular
momentum will change in time. Consequently, the amount of the
rotating material, which was initially available for the torus
formation, may become insufficient at a later stage of the collapsar evolution.

Moreover, the spin of the black hole will be changed by
accretion. Whether the black hole can achieve a spin parameter close
to the maximum one, depends on the properties of the 
accreted mass. While a large spin
 ($a \sim 0.9$) is thought to be a necessary condition 
for the jet launching (Blandford \& Znajek 1977), 
it may happen that not enough specific angular momentum
is being transferred to the black hole as its mass increases.

Another  challenge for the collapsar model is due to the
effects of stellar wind, which removes from the Wolf-Rayet stars a 
large fraction of angular momentum (Langer 1998).
However, the winds of massive stars are relatively weaker in 
the low metallicity environment
(Abbott et al. 1982),
and possibly the GRB progenitor stars can rotate faster 
than an average W-R star
(Vink 2007). 

Here we address the question of whether the collapsing star envelope
contains enough specific angular momentum
in order to support the
formation of the torus. Furthermore, it will be
interesting to study the problem of spin up and spin down
of the newly born black hole, and we shall consider this in a follow up paper. 
These two are the key properties needed to launch 
the GRB jet for
an extended period of time.
Because the angular momentum distribution in the Wolf-Rayet stars
is unknown, we may pose this question also in a different way: 
we want to check how much angular momentum has to be
initially present in the stellar envelope in order to
allow for a long GRB production.

In Section \ref{sec:model}, we describe the model of the initail conditions 
and evolution of the collapsing star, adopting various prescriptions for the angular 
momentum distribution and different scenarios of the accretion process.
In Section \ref{sec:results}, we present the results for the mass accreted onto the black hole
in total and through the torus. We also estimate the 
resulting GRB durations, first in case of a constant $\dot m$ and then
by explicitly calculating the free fall velocity of the gas 
accreting in the torus.
 Finally in Section \ref{sec:diss},
 we discuss the 
resulting duration time of GRBs, as a function of the distribution of the
specific angular momentum in the progenitor star.

\section{Model}
\label{sec:model}

\begin{figure}
\epsscale{.80}
\plotone{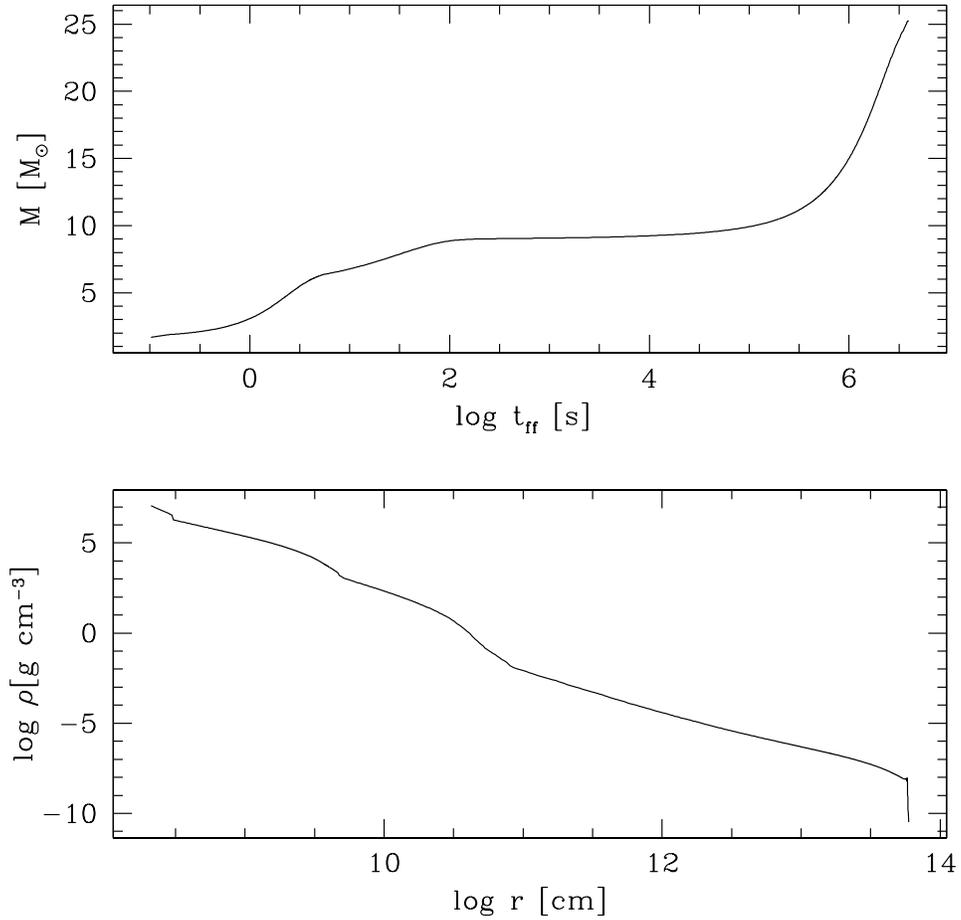}
\caption{The density and mass profiles in the pre-supernova model. The data are taken from 
Woosley \& Weaver (1995), model No. S251S7B@14233.
The x-axis on the upper panel shows the free fall timescale corresponding to the radius shown on the lower panel's x-axis.}
\label{fig:ror}
\end{figure}

In the initial conditions, we use
the spherically symmetric model of the  25 $M_{\odot}$ pre-supernova
(Woosley \& Weaver 1995).
The same model was used by Proga et al. (2003) in their MHD simulation
of the collapsar.
Figure \ref{fig:ror} shows the density profile and the mass enclosed inside a given radius.
The Figure also shows the free fall timescale onto the enclosed mass, 
corresponding to the radius.

The angular momentum within a star or rotating torus may depend on
 radius (see e.g.  Woosley 1995, Jaroszy\'nski 1996, Daigne \&
 Mochkovitch 1997
for various prescriptions). 
Here we parameterize this distribution to be either a function of
the polar angle $\theta$ (simplest case; models {\bf A} and {\bf B}), 
or a function of both
 radius $r$ and $\theta$ (models {\bf C} and {\bf D}).

First, we assume the specific angular momentum to depend only on the
polar angle:
\begin{equation}
l_{\rm spec} = l_0 f(\theta).
\end{equation}
We constitute two different functions:
\begin{equation}
 f(\theta) = 1- |\cos \theta|   ~~ {\rm (model ~ {\bf A})}
\label{eq:ft1}
\end{equation}
\begin{equation}
f(\theta) = \sin^{2}\theta ~~ {\rm (model ~ {\bf B})}
\label{eq:ft2}
\end{equation}
The rotation velocity is therefore given by:
\begin{equation}
v_{\varphi} = {l_{0} \over r \sin \theta} f(\theta)
\end{equation}
The normalization of this dependence is defined with respect to the
critical specific angular momentum for the seed black hole:
\begin{equation}
l_{0} = x l_{\rm crit}(M^{0}_{\rm BH}) = x \times 3.54 \times 10^{16} {M [M_{\odot}]
  \over 2} ~~{\rm cm^{2}~s^{-1}}
\end{equation}
where $R_{\rm g} = 2GM^{0}_{\rm BH}/c^{2}$ 
is the Schwarzschild
radius (non-rotating black hole).

Second, we assume that the specific angular momentum will depend on the
polar angle, as well as on the radius in the envelope, as:
\begin{equation}
l_{\rm spec} = l_{0} g(r)f(\theta),
\end{equation}
 We adopt the following functions:
\begin{equation}
l_{\rm spec} = x l_{\rm crit} ({r \over r_{\rm core}})^{2} \sin^{2}
\theta   ~~ {\rm (model ~{\bf C})}
\label{eq:ft3}
\end{equation}
\begin{equation}
l_{\rm spec} = x \sqrt{8 G M_{\rm core} r} \sin^{2}\theta ~~ {\rm
  (model ~{\bf D})}
\label{eq:ft4}
\end{equation}
The above model {\bf C} corresponds to the rotation with a constant
angular velocity $\Omega$, 
while the model {\bf D} corresponds to a constant ratio 
between the centrifugal and gravitational forces. 
Note that the strong increase of $l_{\rm spec}$ with radius 
will lead to a very fast rotation at large radii.
Therefore, a cut off may be required at some maximum value, $l_{\rm
  max}$ (see below).

The normalization of all the models is chosen such that the specific angular momentum is
always equal to the critical value at $\theta = 90^{\circ}$, and at $r=r_{\rm core}$ if the 
model depends on radius. 
In Section \ref{sec:results}, we present the results of our calculations 
considering a range of initial values of $x$.

Initially, the mass of the black hole is given by the mass of the iron
core
of the star:
\begin{equation}
M_{BH}^{0} = M_{\rm core} = 4 \pi \int_{0}^{r_{\rm core}} \rho r^{2} dr.
\end{equation}
For a given $x$, a certain 
fraction mass of the collapsar envelope, $M^{0}_{1}$,
carries a specific angular momentum smaller than critical $l_{\rm
  crit}^{0} \equiv l_{\rm  crit}(M_{\rm BH}^{0})$:
\begin{equation}
M^{0}_{1} = 2 \pi \int_{r_{\rm core}}^{r_{\rm max}}
\int_{0}^{\pi} \rho_{1} r^{2} \sin \theta d\theta dr
\end{equation}
where $\rho_{1} \equiv \rho (r,\theta)|_{l<l^{0}_{\rm crit}}$ 
is the density in the envelope where 
the specific angular momentum is smaller than critical. 
Here, the radius $r_{\rm max}$ is the size of the star.
Correspondingly, by $M^{0}_{2}$ we denote the fraction of the envelope
  mass that carries the specific angular momentum larger or equal to the critical, with 
$\rho_{2} \equiv \rho (r,\theta)|_{l \ge l^{0}_{\rm crit}}$,
and 
the total envelope mass is $M^{0}_{\rm env} = M^{0}_{1} + M^{0}_{2}$.
Only the mass $M^{0}_{2}$ can form the torus around the black hole of
the mass $M^{0}_{\rm BH}$. 

The above relations set up the initial conditions for the torus
formation in the collapsar, and $l_{\rm crit}$ is defined by the mass
of the iron core, $M_{\rm core}$. However,
as the collapse proceeds, the mass of the black hole will increase and 
the critical specific angular momentum will be a function of the increasing
mass: $l_{\rm crit}(M_{\rm BH})$.
The main point of this work is to compute the mass of the progenitor
with $l>l_{\rm crit}$, taking into account this effect.
Below, we redefine $\rho_{1}$ and $\rho_{2}$, so that the
$l_{\rm crit} - M_{\rm BH}$ relation is taken into account.

To compute the mass of the envelope part that has high enough $l$ to
form a torus around a given BH, and to
estimate the time duration of the GRB powered by accretion, 
we need to know the mass of this black hole.
A current $M_{\rm BH}$ depends on the mass of the mass of the seed
black hole and the accretion scenario.
We approximate this scenario in a following way. We assume that
accretion is nonhomologous and BH grows by accreting mass $\Delta
m^{\rm k}$, which is a function of the mass of a shell between the
radius $r_{\rm k}$ and $r_{\rm k}+\Delta r_{\rm k}$ (see e.g. Lin \&
Pringle 1990). 
Formally, we perform the calculations of  $M_{\rm BH}$ and
$\Delta m^{\rm k}$ iteratively:

\begin{equation}
M_{\rm BH}^{\rm k} = M_{\rm BH}^{\rm k-1}+\Delta m^{\rm k}
\end{equation} 
where the increment of mass of the black hole is :
\begin{equation}
\Delta m^{\rm k} = 2 \pi \int_{r_{\rm k}}^{r_{\rm k}+\Delta r_{\rm k}}
\int_{0}^{\pi} \bar{\rho} r^{2} \sin \theta d\theta dr
\end{equation}
Here $\bar{\rho}$ depends on the accretion scenario (see below)
and contains the information of the specific angular momentum distribution.
The above two equations define an iterative procedure due to the
nonlinear dependence of $\bar{\rho}$ on $M_{\rm BH}$.
We start from the radius $r_{0} = r_{\rm core}$, i.e. that of an iron core.

We distinguish here three possible accretion scenarios: \\
(a) the accretion onto black hole proceeds at the same rate both from
the torus and from the gas close to the poles, with $l<l^{\rm k}_{\rm crit}$, i.e.
 $\bar{\rho} \equiv \rho$ (and does not depend on $\theta$); \\
(b) the
envelope material with $l<l^{\rm k}_{\rm crit}$ falls on the black hole first.
Thus, until the polar funnel is evacuated completely, only this gas contributes to the
black hole mass, i.e. $\bar{\rho} \equiv \rho_{1}$. After that, the material
with $l>l^{\rm k}_{\rm crit}$ accretes, and $\bar{\rho} \equiv \rho_{2}$; \\
(c) the accretion proceeds only through the torus, and only this
material contributes to the black hole growth
i.e. $\bar{\rho} \equiv \rho_{2}$. In this case the rest of the envelope 
material is  kept aside until the torus is accreted.

\begin{figure}
\epsscale{.80}
\plotone{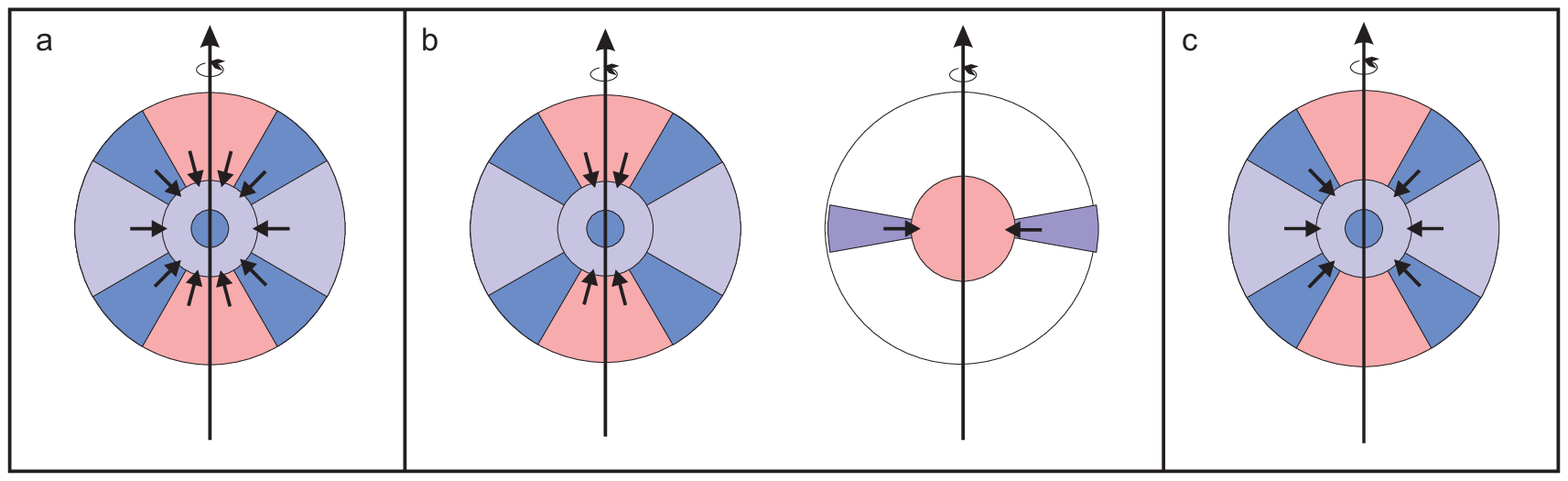}
\caption{The scheme of accretion scenarios. The red color 
indicates the material with $l<l_{\rm crit}$. The 
blue colors indicate the material with $l>l_{\rm crit}$:
darker for smaller black hole mass, and lighter for larger black hole mass.
 Arrows indicate, which material is accreting and contributes to the black hole growth.}
\label{fig:scheme}
\end{figure}

The densities $\rho_{1}$ and
$\rho_{2}$, defined above, 
depend on $l_{\rm crit}^{\rm k} \equiv l_{\rm crit}(M^{\rm k}_{\rm BH})$. 
The above accretion scenarios are illustrated in the Figure 
\ref{fig:scheme}. The panel (a) shows the scenario of a uniform accretion,
in which the whole envelope material falls into black hole, 
regardless of its specific angular momentum. The red color marks
the material with $l<l^{\rm k}_{\rm crit}$.
The blue colors mark the material with $l>l^{\rm k}_{\rm crit}$, and when the black hole is 
small, this condition is satisfied for a larger range of $\theta$ (dark blue).
When the black hole increases, the material with  $l>l^{\rm k}_{\rm crit}$ occupies narrower
 $\theta$ range (light blue).
 The panel (b) shows the scenario with two steps: first the material
 with  $l<l^{\rm k}_{\rm crit}$
 accretes onto the black hole, increasing its mass; after this material is
 exhausted, the material with $l>l^{\rm k}_{\rm crit}$ starts accreting. Because the black hole
 mass has already increased, material with large $l$ is concentrated very close to the equator.
 The panel (c) shows the scenario in which only the material with 
 $l>l^{\rm k}_{\rm crit}$ accretes.

In scenario {\it a} the mass accretion rate does not depend on
the specific angular momentum. This is a very special and
simple scenario. In reality, the accretion rate can depend
on the specific angular momentum.
For example, if an accreting torus produces a very powerful
outflow, the weakly rotating polar material could be expelled
and never reach the black hole (scenario {\it c}). This would also be a
special and extreme situation. It is more likely
that both the polar and disk material accrete but at different
rates. However, it is unclear what these rates are
and detailed simulations of MHD flows show that
the rates can depend on time. For example, there are periods
of time when
the polar material accretes faster than the disk material
and vice versa (e.g. Proga \& Begelman 2003). To bracket this more realistic
situation, we consider here another extreme and rather
artificial scenario {\it b}, which corresponds
to an somewhat 'reversed' scenario {\it c}. In this scenario, 
initially the torus accretion rate is  zero and accretion is dominated
by the polar material.
Only after the polar mateial is exhausted,
the torus accretion starts. We note that although
this scenario is quite extreme, it may be relevant if
jets in GRBs must be very clean and light because
in this scenario jets will be moving in the 'empty' polar funnels.

Due to the increasing mass of the black hole, the critical angular
momentum also increases, and as a result less and less material can
satisfy the condition for the torus formation ($l > l^{\rm k}_{\rm crit}$).
We stop the calculations, when there is no material with 
$l>l^{\rm k}_{\rm crit}$, 
i.e. able to form the torus:
\begin{equation}
w_{\rm k} = {M_{2}^{\rm k} \over M_{\rm env}^{\rm k}} = 
{ 2 \pi \int^{r_{\rm max}}_{r_{\rm k+\Delta r}}
\int_{0}^{\pi} \rho_{2} r^{2} \sin \theta d\theta dr
\over 
4 \pi \int^{r_{\rm max}}_{r_{\rm k+\Delta r}}
 \rho r^{2} dr
} = 0.
\label{eq:kmax}
\end{equation} 
 Alternatively,
the iterations may be stopped earlier, for example if we impose a physical limit 
based
on the free fall timescale or the accretion rate, to be adequate to 
power the prompt GRB phase.

The duration of the GRB could be estimated as the ratio between
 the mass accreted through the torus, and the accretion rate $\dot m$:
\begin{equation}
M^{\rm torus}_{\rm accr} = \sum_{\rm k=1}^{k_{\rm max}} M_{2}^{\rm k}
\end{equation}
\begin{equation}
\Delta t_{\rm GRB} = {M_{\rm accr}^{\rm torus} \over \dot m} 
\end{equation}
where the number $k_{\rm max}$ is defined by the Equation \ref{eq:kmax}.

Note that we assume here the GRB prompt emission is equal to the
duration of the torus replenishment.
In principle, $\dot m$ may depend on time.
 Here we take two approaches. First,  
for simplicity, we assume a constant accretion rate
of a moderate value ($\dot m = 0.01-1.0$ M$_{\odot}$ s$^{-1}$, 
see e.g. Popham, Woosley \& Fryer 1999; Janiuk et al. 2004).
Second, in more detailed calculations we determine the instantaneous
accretion rate during the iterations, determined by the free 
fall velocity of gas in the torus.

\section{Results}
\label{sec:results}

\subsection{Models with the specific angular momentum dependent only on $\theta$}
\label{sec:theta}

\begin{figure}
\epsscale{.80}
\plotone{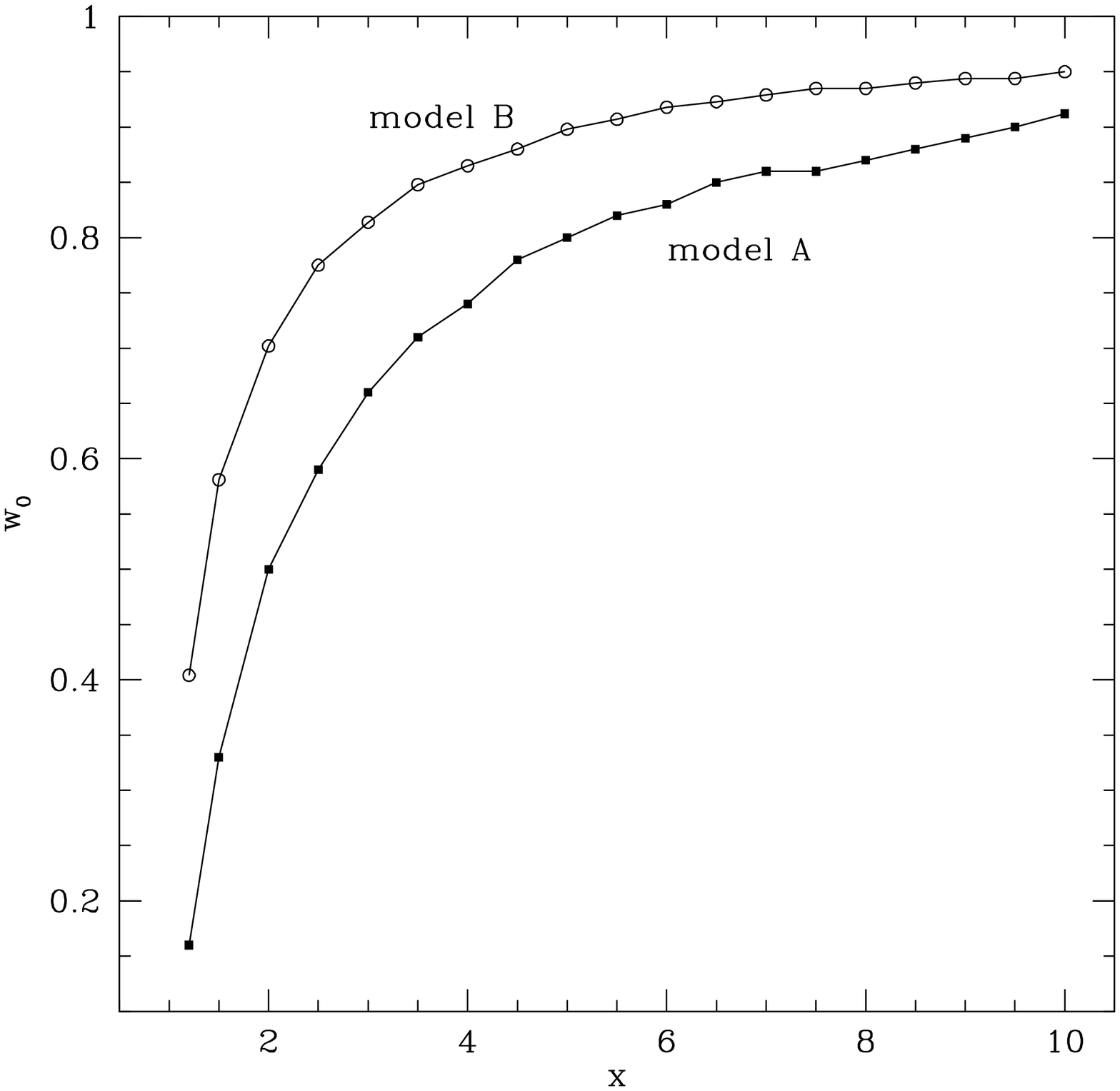}
\caption{The initial mass fraction of material with the angular
  momentum $l>l_{\rm crit}$, as a function of the
initial normalization of the specific angular momentum distribution,
for model {\bf A} (solid squares) and model {\bf B} (open circles) of the
  distribution function $f(\theta)$.
}
\label{fig:fig3}
\end{figure}

The Figure \ref{fig:fig3} shows the initial fraction of the envelope mass 
which contains large enough angular
momentum to form the
rotating torus, $w_{0} \equiv M^{0}_{2}/M^{0}_{\rm env}$ 
(see Eq. \ref{eq:kmax})
 for models {\bf A} and {\bf B}.
For instance, for the adopted function
$f(\theta)$ given by Eq. \ref{eq:ft1} (model {\bf A}),
 and for the initial angular
momentum normalization of $x=1.15$, we obtain  $w_{0}=0.13$. 
This means that only 13\% of the total mass of the
envelope will initially be able to form the torus,
 while the remaining 87\% of the
envelope mass will fall radially towards black hole.
On the other hand, for $x>5$,
more than 75\% of the envelope mass will be rotating fast enough to
contribute to the torus formation.
The model {\bf B} gives systematically larger values of $w_{0}$
and for $x=1.15$, we have $w_{0}=0.36$, while for $x> 5$ we have 
more than 85\% of the envelope mass able to form the torus.

As we show below, these are only the upper limits for the mass that could be 
accreted through the torus, and drive the GRB duration. These values will be much smaller, 
when we calculate the collapsar evolution with increasing 
$l^{\rm k}_{\rm crit}$ instead of $l^{\rm 0}_{\rm crit}$.

\begin{figure}
\epsscale{.80}
\plotone{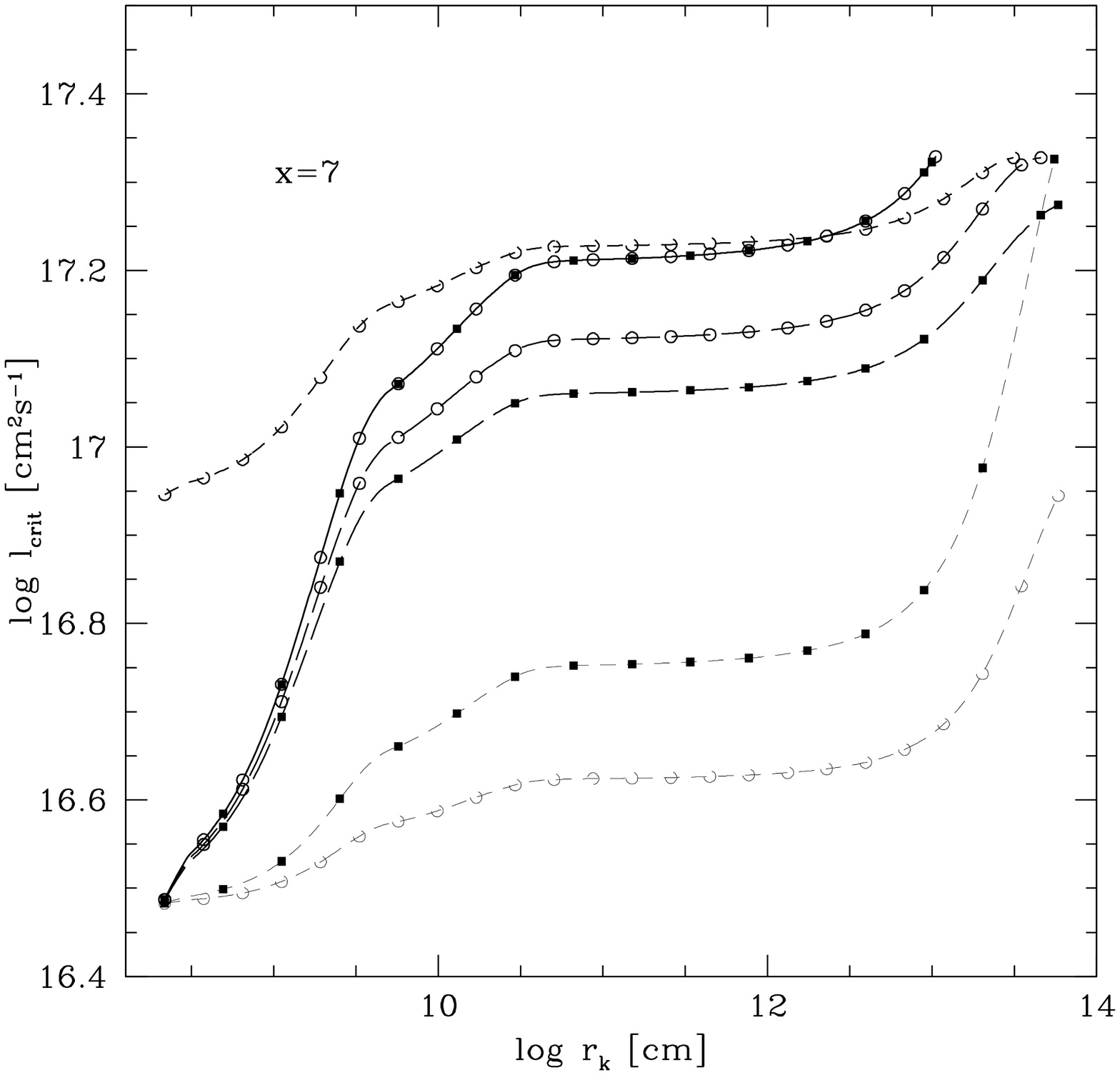}
\caption{The critical specific angular momentum
as a function of radius within which the envelope mass collapsed 
onto BH. The results are shown for one exemplary value of the initial
normalization parameter, $x=l_{0}/l_{crit} = 7$.
The models of the distribution function $f(\theta)$ are: 
{\bf A} (solid squares) and {\bf B} (open circles),
and accretion scenarios are: {\it a} (solid line), {\it b} (short dashed line)
and {\it c} (long dashed line). The thin line for model {\bf B}{\it b} 
represents the results from the phase 1 of accretion (through the
poles), while the thicker line {\bf B}{\it b} represents the results
from the phase 2 (torus accretion).
Note, that for the model {\bf A}{\it b} there is only a thin line, and 
no phase 2, because for $x=7$ the torus does not form. 
Note also, that the solid lines (i.e. models {\bf A}{\it
a} and {\bf B}{\it a} overlap.}
\label{fig:lc}
\end{figure}

The Figure \ref{fig:lc} shows $l^{\rm k}_{\rm crit}$, i.e. $l_{\rm crit}$ as a function of the
current radius $r_{\rm k}$, which is the inner radius of the collapsing
envelope in a subsequent step $k$.
The figure thus shows how the critical specific angular momentum changes 
with time during the collapse, for an exemplary value of $x=7$. 

The $l_{\rm crit}$ rises with time, 
as the black hole accretes mass from the envelope, and corresponds to
a changing black hole mass. The 
most steep rise is for the uniform accretion scenario {\it a}, and in this
case by definition the result does not depend on the adopted distribution function
 for the specific angular momentum, $f(\theta)$. Therefore both curves marked by a solid line
 overlap.
Also, in scenario {\it a} the plotted
curves do not depend on $x$, as well as neither on the slope nor on the
location of the curve. The latter influences only the maximum of
this curve, as for larger $x$ we have more material available to
form the torus. In particular, for $x=7$, the two overlapping 
curves shown in the figure end at $r\sim 10^{13}$ cm.

For the scenario {\it c}, i.e. accretion of gas with $l>l^{\rm k}_{\rm crit}$,
$l_{\rm crit}$ rises less steeply with $r_{\rm k}$ than in scenario
{\it a}, because
now less material is contributing to the black hole mass. 
In this case $f(\theta)$ 
affects the results, and
model {\bf A}
gives systematically smaller values of 
$l_{\rm crit}$ than model {\bf B}. 

For the scenario {\it b}, the result is very sensitive to $x$,
and we can have either one or two phases of accretion: only the 
polar inflow or first the polar inflow
and then torus accretion.
The value of $x=7$ was chosen, because in model {\bf A} still no torus
is able to form,
and we have only phase 1,
while in model {\bf B} this value of $x$ is already large enough and
the phase 2 occurs.

For phase 1 in scenario {\it b} (marked by the thinner lines in the figure), 
i.e. the material with $l<l^{\rm k}_{\rm crit}$ is accreting, the 
dependence on
$f(\theta)$ is the following:
model {\bf A} adds more mass to the black hole and therefore it 
leads to the larger values of $l_{\rm crit}$ than model {\bf B}.
For phase 2 of scenario {\it b} (present only
for model {\bf B} and marked by the thick line in the figure),
the evolution starts from the last $l^{\rm k}_{\rm crit}$ achieved in the end
of phase 1. 
Then $l^{\rm k}_{\rm crit}$ increases, and ultimately reaches the final 
solution of models {\bf B}{\it c} and {\bf A}{\it b},
because this $l_{\rm crit}$ corresponds to the black hole mass 
that has increased maximally: either only through a torus, or first 
through the through the polar funnels and then through the torus accretion.

All the curves in Figure \ref{fig:lc} exhibit a characteristic
evolution of their slopes, tracing the density distribution in
the progenitor star (see the top panel of Figure 
\ref{fig:ror}). First, the fast rise is due to accretion of
the most dense inner shells of the stellar envelope. Then, the slope
flattens,  as the density in the envelope decreases and the mass does not
grow very fast. 
In the end, the slope of 
$l^{\rm k}_{\rm crit} \equiv l_{\rm crit}(r_{\rm k}) \equiv 
l_{\rm crit}(M_{\rm BH})$ rises again, 
due to larger
volume of the shells, but this rise is depending on
the adopted scenario. In scenario {\it c} the sequence is the following:
 increase of $l^{\rm k}_{\rm crit}$ $\rightarrow$  more accretion 
$\rightarrow$ larger increase of $l^{\rm k}_{\rm crit}$ $\rightarrow$ less
accretion. In the phase 1 of 
scenario {\it b}, such an equilibrium is not established, because
 the accretion onto black hole proceeds through the polar funnels,
i.e. using the material with $l<l^{\rm k}_{\rm crit}$, so we
have: increase  of $l^{\rm k}_{\rm crit}$ $\rightarrow$ more accretion
$\rightarrow$ further increase  of $l^{\rm k}_{\rm crit}$, and for
large radii $r_{\rm k}$ the slope of
the curves shown in Fig. \ref{fig:lc} in this scenario is much steeper. 
The phase 2 of 
scenario {\it b} results in the changes of the slope of $l^{\rm
  k}_{\rm crit}$
 similar to other scenarios, {\it a} and {\it b} (provided that the
 phase 2. occurs). 

\begin{figure}
\epsscale{.80}
\plotone{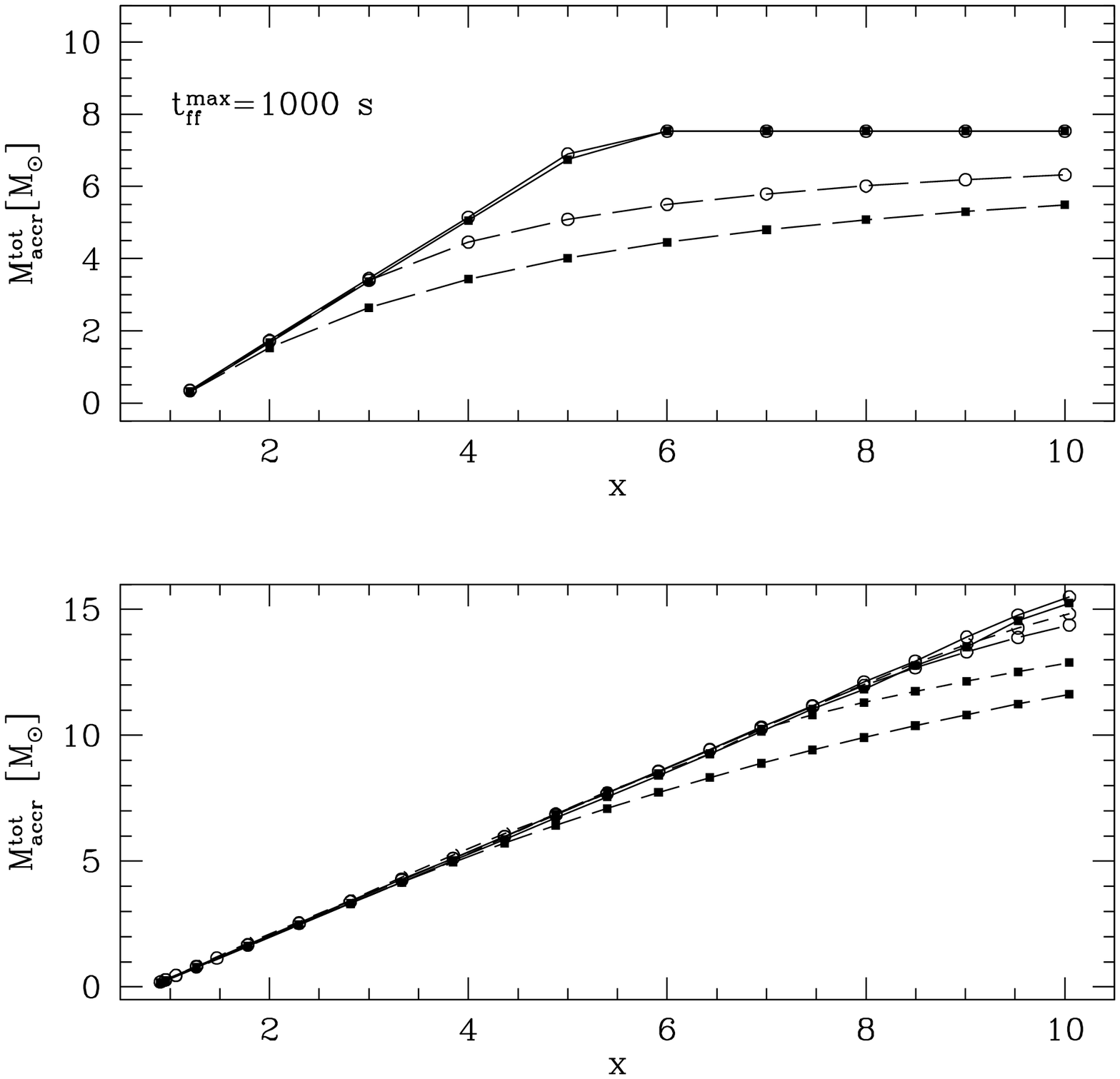}
\caption{The total  mass accreted onto a black hole, 
as a function of the
initial normalization of the specific angular momentum distribution.
The models of the distribution function $f(\theta)$ are: 
{\bf A} (solid squares) and {\bf B} (open circles),
and accretion scenarios are: {\it a} (solid line), {\it b} (short dashed line)
and {\it c} (long dashed line).
The upper panel shows the case when the mass accretion is limited by a 
maximum free fall time of 1000 seconds, while the lower panel shows the
results for no limiting $t_ff$ (cf. Fig. \ref{fig:ror}).}
\label{fig:fig2}
\end{figure}

\begin{figure}
\epsscale{.80}
\plotone{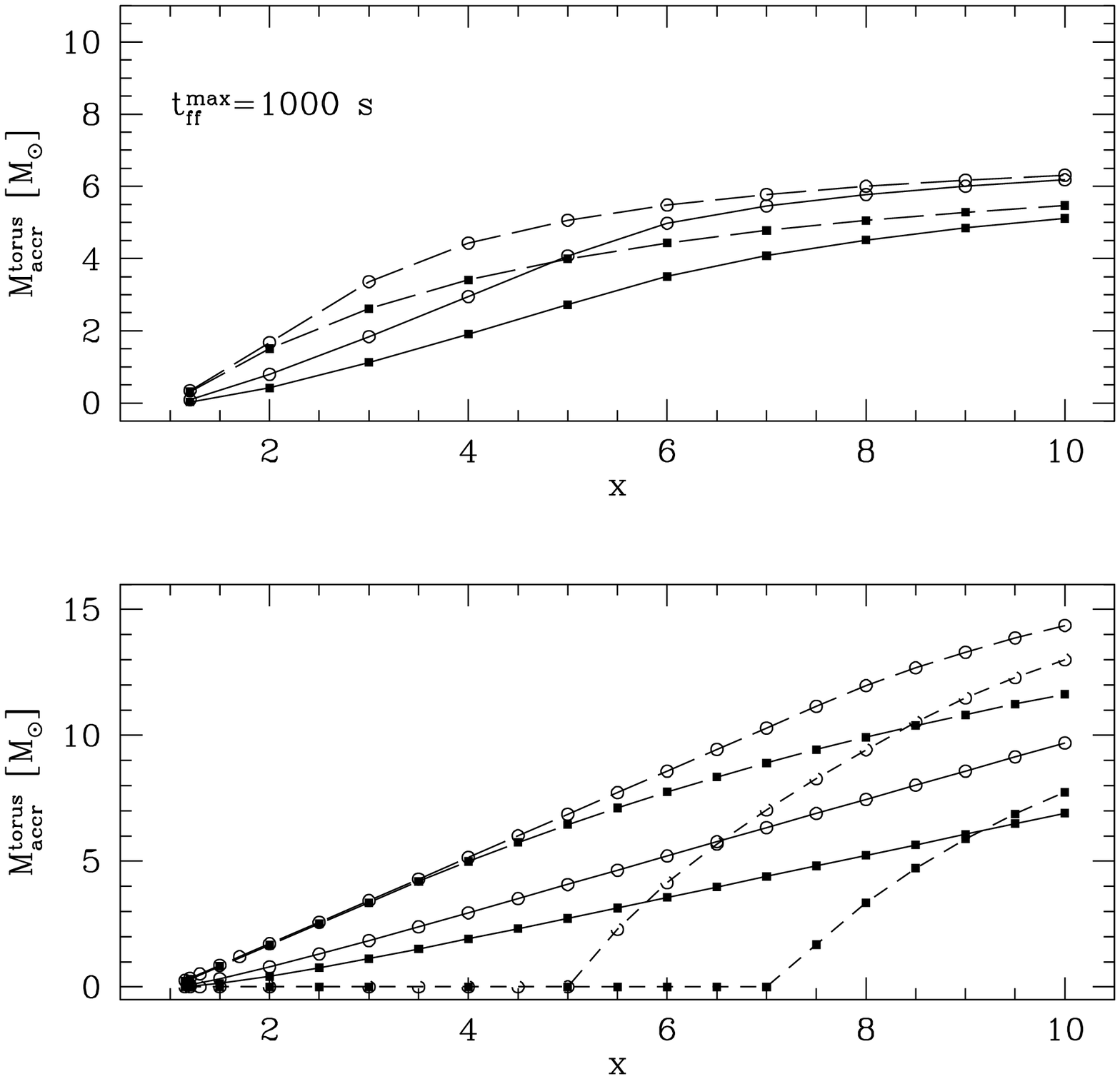}
\caption{The accreted mass with $l>l_{\rm crit}$, 
as a function of the
initial normalization of the specific angular momentum distribution.
The models of the distribution function $f(\theta)$ are: 
A (solid squares) and {\bf B} (open circles),
and accretion scenarios are: {\it a} (solid line), {\it b} (short dashed line)
and {\it c} (long dashed line).
The upper panel shows the case when the mass accretion is limited by a 
maximum free fall time of 1000 seconds, while the lower panel shows the
results for no limiting $t_ff$ (cf. Fig. \ref{fig:ror}).}
\label{fig:fig2a}
\end{figure}

In Figure \ref{fig:fig2}, we show the total mass accreted onto a black
hole,
and in Figure \ref{fig:fig2a} we show the mass accreted through the torus,
both
as a function of $x$. Figure \ref{fig:fig2a} can be also regarded as
showing
 the estimated duration of a GRB, if the
accretion rate is constant.
The results are again presented for the 3 scenarios of accretion: {\it a}, 
{\it b}
and {\it c} 
(marked by solid, short-dashed and long dashed lines, respectively), 
as well as for the two prescriptions for the function
$f(\theta)$, models {\bf A} and {\bf B} 
(marked by squares and circles, respectively).
 The upper panels in Figs. \ref{fig:fig2} and \ref{fig:fig2a}
show the results obtained is case of the maximum limit for the
free fall time set yo $t_{\rm ff}^{\rm max}=1000$ s.

The values of the total accreted mass
are the largest for the scenario of the uniform accretion, {\it a}.
Depending on $x$, 
 the black hole mass is growing until
there is no further material with $l > l^{\rm k}_{\rm crit}$.

In scenario {\it b}, initially we add to the black hole mass 
(thus increasing $l^{\rm k}_{\rm crit}$), 
only the material with $l<l^{\rm k}_{\rm crit}$. For small values of $x$, 
the total accreted mass is the same as for scenario {\it a}, because
the process of accretion lasts in both cases until $M^{\rm k}_{2} = 0$,
i.e. the envelope contains no 
further material with $l>l^{\rm k}_{\rm crit}$.
However, for large $x$ ($\ge 7$ in model {\bf A} and $\ge 5$ in model
{\bf B}),
after the black hole has swallowed the whole funnel with 
$l<l^{\rm k}_{\rm crit}$, there is still some material with large
specific angular momentum and phase 2 occurs.
The material accretes now through the torus, but only as long as it has
$l>l^{\rm k}_{\rm crit}$.  Therefore, for large $x$, 
the total mass which the black hole gains in scenario {\it b} is 
 less than in the scenario {\it a}.

Scenario {\it c} assumes, that the BH accretes only the material 
with $l>l^{\rm k}_{\rm crit}$. Now,
the total accreted mass can be either the same (for small $x$) 
or smaller than in scenario {\it a}.

In Figure \ref{fig:fig2a}, we show the 
accreted mass  which had $l>l^{\rm k}_{\rm crit}$.
This represents the accretion through the torus, and may be
regarded as a direct measure of the GRB duration.
Scenario {\it a} results in a linear scaling of 
$M_{\rm accr}^{\rm torus}$ with $x$:
\begin{equation}
M_{\rm accr}^{\rm torus} =  \alpha x + \beta
\end{equation}
and with a linear fit we obtained
$\alpha \approx 0.83$, $\beta \approx -1.41$ in model {\bf A}, and $\alpha\approx
1.12$, $\beta \approx -1.55$ in model {\bf B}.

Scenario {\it b} predicts that the torus accretion is possible only
for large $x$, while for small $x$ torus will 
not form, as discussed above.
The scenario {\it c}, by definition, 
predicts the torus accretion for any value of $x>1.0$.
Therefore, the amount of material
accreted with $l>l^{\rm k}_{\rm crit}$ is in this scenario
larger than 
in scenario {\it a}, because the black hole mass grows more slowly.
Both scenarios {\it a} and {\it b} result, for large $x$, in a nonlinear
dependence of $M_{\rm accr}^{\rm torus}$ on $x$.

To sum up, in the models {\bf A} and {\bf B}, i.e. if the specific angular
momentum in the progenitor star depends only on the polar angle,
the total mass of material capable of forming the torus can only be a
small fraction of the envelope mass. Depending on the accretion scenario, it
is at most $\sim 3.5 M_{\odot}$,  i.e. 15\% of the envelope mass,
for $l_{0} = 3 l^{0}_{\rm crit} = 10^{17}$ cm$^{2}$s$^{-1}$, and 
between $\sim 7$ and $\sim 15 M_{\odot}$,  i.e. 30\%-65\% of the
envelope mass, for $l_{0} = 10 l^{0}_{\rm crit} = 
3.3 \times 10^{17}$ cm$^{2}$s$^{-1}$.

Note that we could proceed with larger $l_{0}$, but we 
stopped our calculations at  $x=10$,
because larger $x$ would already imply very fast rotation at the
equator. In the present section we did not assume any maximum limit
on the specific angular momentum; this will be taken into account in
the next section.
 Howewer, we considered the effects of the maximum limit for the free
fall timescale. As shown in the Figures, the limit of 
$t_{\rm ff}^{\rm max}=1000$ s plays an important role when $x>5$, 
and in all the models and scenarios
 the dependence of the accreted mass on $x$ is significantly weaker 
than for the case of no limiting $t_{\rm ff}$.
For large $x$, in scenario {\it a} the total mass accreted onto BH is 
constant with $x$ 
and equal to the fraction of the envelope mass enclosed within the radius
$r \approx 1.58 \times 10^{11}$ (see Fig. \ref{fig:ror}).
The mass accreted via torus is smaller than that, and for $x=10$ 
it reaches about 6 solar masses.

\subsection{Models with the specific angular momentum dependent on $r$ and  $\theta$}
\label{sec:radius}

Now, we investigate how the total accreted mass and in consequence the
duration of the GRB will be affected if the angular velocity $\Omega$ 
in the collapsing star  is constant or given by a fixed ratio of the
centrifugal to gravitational force (Equations \ref{eq:ft3} and \ref{eq:ft4},
respectively).

In these two models, {\bf C} and {\bf D}, the specific angular momentum is
a strong function of radius.
Therefore, if we do not impose any limit on the specific angular momentum, 
 $l_{\rm max}$, the outer shells of the star will always have a 
substantial amount of specific angular momentum, larger than any
current critical
value. Consequently, the GRB will continue until the last shell is 
accreted. However, this would imply a very fast rotation of the star in its outer layers.
This would be inconsistent with the progenitor models for GRBs, so
in a more  realistic version of our modeling we impose a
maximum value of the specific angular momentum. This assumption will lead
to a shorter GRB event, as it will end when the increasing black hole
mass will imply $l^{\rm end}_{\rm crit} > l_{\rm max}$.

\begin{figure}
\epsscale{.80}
\plotone{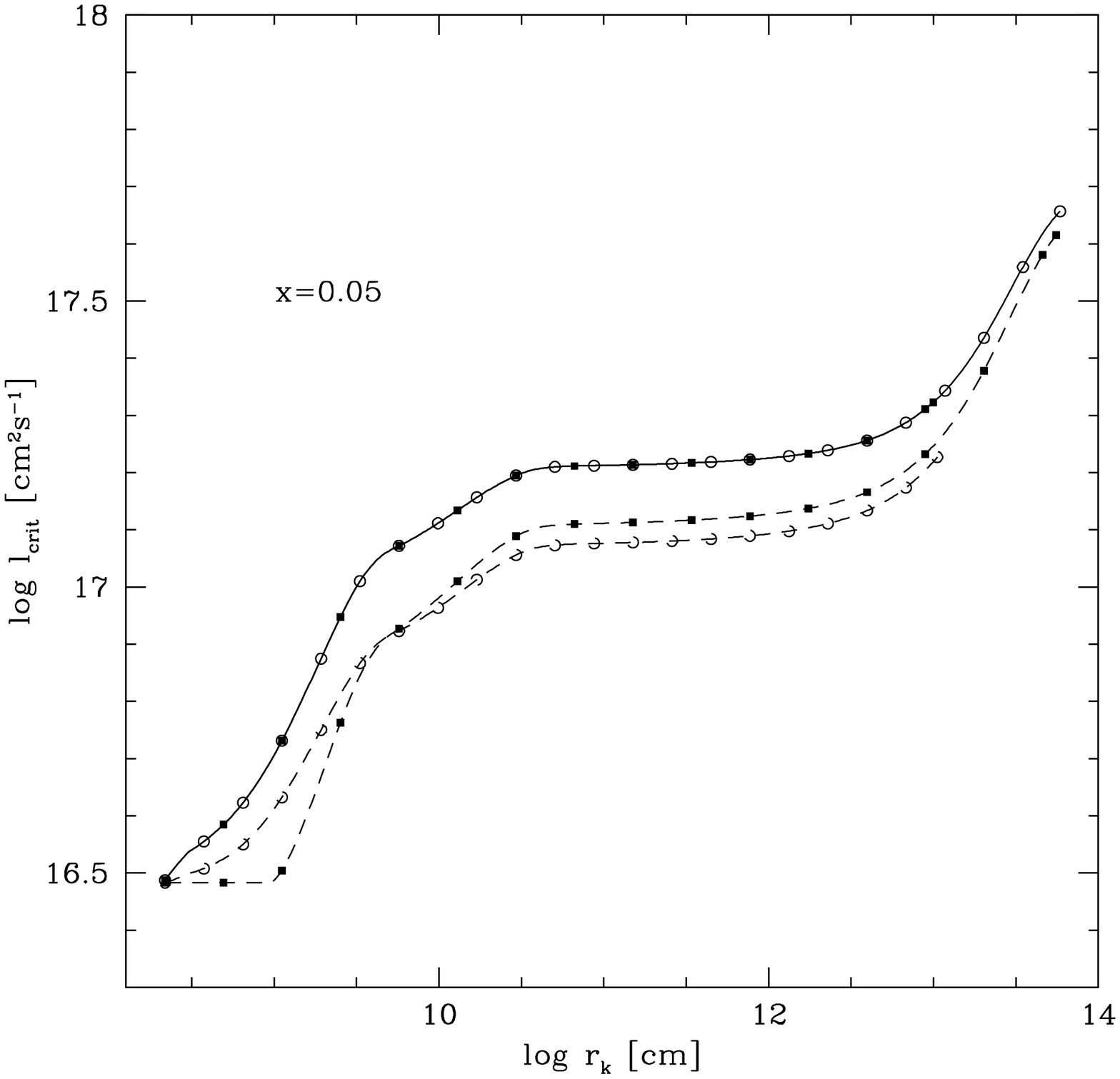}
\caption{The value of critical specific angular momentum
during an iteration, for one exemplary value of the initial
normalization parameter, $x=l_{0}/l_{\rm crit} = 0.05$.
The models of the distribution function $f(r,\theta)$ are: 
{\bf C} (solid squares) and {\bf D} (open circles),
and accretion scenarios are: {\it a} (solid line) and {\it c} (dashed line).
}
\label{fig:figlccd}
\end{figure}

First, we investigate how
the black hole mass, $M^{\rm k}_{\rm BH}$, and critical specific
angular momentum, $l^{\rm k}_{\rm crit}$,
depend on the accretion scenario.
For the scenario {\it a} (i.e. uniform accretion), the black hole is
fed by the whole envelope regardless of the local specific angular momentum value.
The result is the same as in the cases explored in Section
\ref{sec:theta}: 
the total accreted mass does not depend neither on 
the distribution function ({\bf C} or {\bf D}) nor on the normalization ($x$).
In the Figure \ref{fig:figlccd},
 we show how the critical specific angular momentum
increases when the subsequent envelope shells are accreting (i.e.
as a function of radius). The solid lines in this figure
overlap, and basically the curve is the same as in the
figure \ref{fig:lc} (for models {\bf A} and {\bf B}), 
the only difference being that the maximum value reached in models
{\bf C} and {\bf D} can be larger (specifically, for
 $\log l_{\rm max} \ge 17.3$).

\begin{figure}
\epsscale{.80}
\plotone{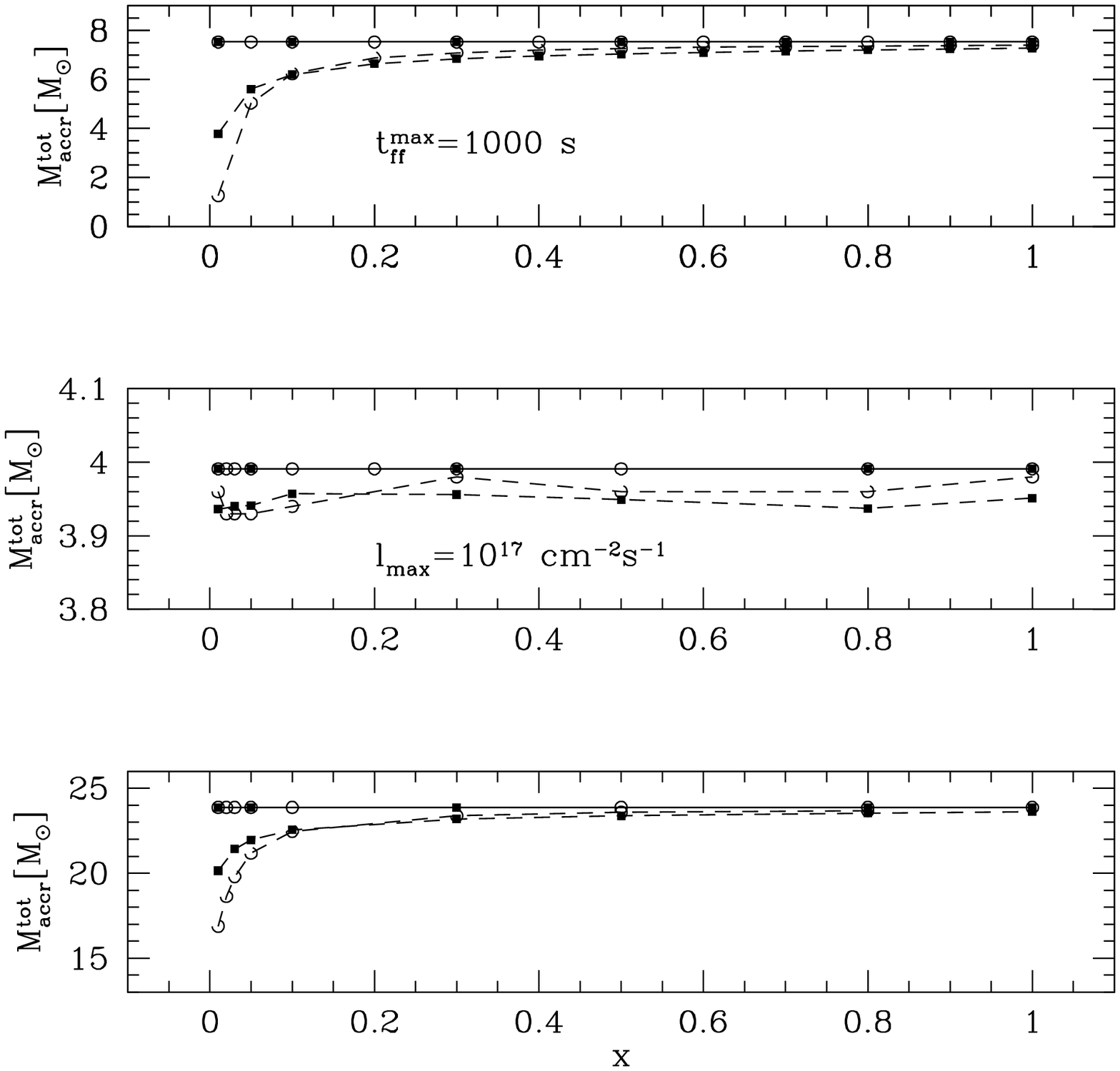}
\caption{
The total accreted mass
  as a function of the
initial normalization of the specific angular momentum distribution.
The models of the distribution function $f(r)g(\theta)$ are: 
{\bf C} (solid squares) and {\bf D} (open circles),
and accretion scenarios are: a (solid line) and c (dashed line).
The upper panel shows the case when the mass accretion is limited by a 
maximum free fall time of 1000 seconds,
the middle panel shows the case of the specific angular momentum cut off to
$l_{\rm max} = 10^{17}$ cm$^{2}$s$^{-1}$, while the bottom panel shows the case of
no free fall time limit and no specific angular momentum cut off.
Note, that the solid lines for models {\bf C} and {\bf D} overlap.
}
\label{fig:figmcd}
\end{figure}

The amount of the total 
accreted mass in this case is constant
(Figure \ref{fig:figmcd}) and the value depends only on $l_{\rm max}$:
\begin{equation}
M_{\rm accr} = {l_{\rm max}c \over 4 G} - M_{\rm core}.
\label{eq:maccr}  
\end{equation}
For our model of the star and $l_{\rm max} = 10^{17}$ cm$^{2}$s$^{-1}$
 this gives
$M_{\rm accr} = 3.99 M_{\odot}$. 
If there is no cutoff
of $l_{\rm max}$ then simply the total envelope mass is accreted,
 23.9 $M_{\odot}$
(cf. the bottom panel of Fig. \ref{fig:figmcd}).

The situation becomes more complicated when we adopt the scenario {\it
  c} (accretion of material with $l> l^{\rm k}_{\rm crit}$).
The accreted mass in this scenario depends both on the distribution and on
the normalization of the specific angular momentum in the 
pre collapse star.

For small $x$, the accreted mass is small, because 
the process ends when $l<l^{\rm k}_{\rm crit}$ everywhere in the
envelope and $M^{\rm k}_{2}=0$.
The total mass accreted on black hole
is sensitive to the model distribution function.
In particular, the fact that this function strongly depends on the
radius means that the inner shells contain mostly material with 
$l \ll l^{\rm k}_{\rm crit}$. Thus, only the more distant shells contribute 
to the mass of the black hole 
(see Fig. \ref{fig:figlccd}, dashed lines), 
and the particular value of $r_{\rm k}$ for
which the black hole mass and $l^{\rm k}_{\rm crit}$ start rising 
depends on $x$. Note that Fig. \ref{fig:figlccd} shows only the
results for $x=0.05$ (arbitrarily chosen value).

For large $x$, the accreted mass  will
asymptotically reach the result for the scenario {\it a}
(see Figure \ref{fig:figmcd}, bottom panel), because
if $x = 1$ then the whole envelope material satisfies
$l \ge l_{\rm crit}$. 
Therefore the
$l_{\rm crit}(M_{\rm BH})$ functions seen in the Figure 
\ref{fig:figlccd} will eventually overlap if
$x$ is close to 1.0.
Therefore in the Figure \ref{fig:figmcd} we show
only the results for $x \le 1.0$ because these are the most
interesting:
for larger $x$ the mass accreted both through the torus and in total,
approaches a constant value.
 
The smallest amount of the total accreted mass 
is obtained when we impose a
cutoff limit on the specific angular momentum,  $l_{\rm max}$. This is shown
in  Fig. \ref{fig:figmcd} (middle panel) for the value
of $l_{\rm max}=10^{17}$ cm$^{2}$s$^{-1}$.
The total accreted mass is in this model $M_{\rm accr}^{\rm tot}\ll M_{\rm env}$, 
and very weakly depends on $x$.

The value of the  $l_{\rm max}$ cut off has to be chosen carefully,
because if
$l_{\rm max} \ge {4G \over c}M_{\rm env} = 4.23\times 10^{17}$ 
cm$^{2}$s$^{-1}$, then
the accreted mass would be equal to the envelope mass (and equal to
that obtained with the uniform accretion scenario), for $x=1$.
Any value of $l_{\rm max}$ larger than the above value will
not affect the results.

The chosen form of the specific angular momentum distribution 
(models {\bf C} or {\bf D}) only slightly
affect the results. 
For $x \sim 0.01$, the model {\bf C} gives larger value of accreted
mass, while for $x \ge 0.1$, 
the model {\bf D} leads to somewhat larger $M_{\rm acc}^{\rm
  tot}$. However, the results in this case are also affected by the numerical
issues because of the very small number of steps after which the
calculation is finished.

The accreted mass will be zero, and the 
 burst will not occur, if the normalization $x$ is very small.
The minimum value can be estimated for {\bf C} and {\bf D} models separately,
if we take the specific angular momentum to be everywhere smaller than critical:
\begin{equation}
x_{\rm min}^{\rm C} = ({r_{\rm core} \over r_{\rm max}})^{2} = 
1.22\times 10^{-11}
\end{equation}
and
\begin{equation}
x_{\rm min}^{\rm D} = {4 \over c}\sqrt{G M_{\rm core} \over r_{\rm max}} = 2.6\times 10^{-4}.
\end{equation}

\begin{figure}
\epsscale{.80}
\plotone{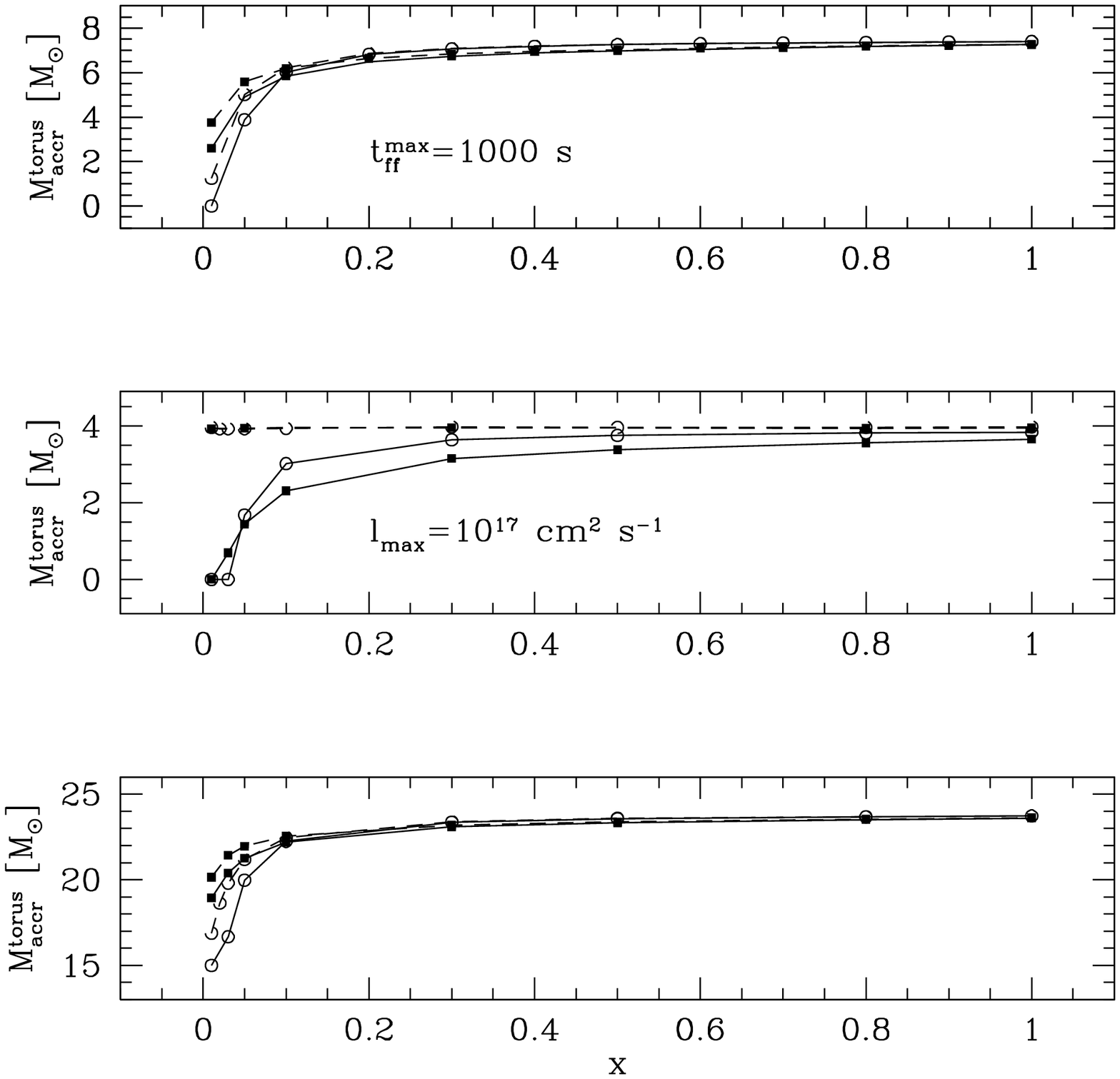}
\caption{
The mass accreted with $l>l_{\rm crit}$
  as a function of the
initial normalization of the specific angular momentum distribution.
The models of the distribution function $f(r)g(\theta)$ are: 
{\bf C} (solid squares) and {\bf D} (open circles),
and accretion scenarios are: a (solid line) and c (dashed line).
The upper panel shows the case when the mass accretion is limited by a 
maximum free fall time of 1000 seconds,
the middle panel shows the case of the specific angular momentum cut off to
$l_{\rm max} = 10^{17}$ cm$^{2}$s$^{-1}$, 
while the bottom panel shows the case of no free fall time limit and
no specific angular momentum cut off.
}
\label{fig:mcdtorus}
\end{figure}

Now, we can 
estimate the duration of GRB by means of the mass accreted onto
the black hole with $l<l^{\rm k}_{\rm crit}$, i.e. via the torus.
In scenario {\it c}, this is, by definition,
the same as the total mass accreted. As can be seen from the bottom
panel in the
Figure \ref{fig:mcdtorus},  $M_{\rm acc}^{\rm torus} \ge 20 M_{\odot}$, and GRB
duration on the order of $\sim$ 40 seconds, is possible only with the 
model with no cut off on $l_{\rm max}$ and $x > 0.1$.
For both models {\bf C} and {\bf D} 
the uniform accretion scenario {\it a} gives slightly 
less amount of mass accreted
through the torus, but the difference is visible only for $x < 0.1$.

The more physical models with the angular cut off to $l_{\rm max} =
10^{17}$ cm$^{2}$ s$^{-1}$ always give less than  $\sim 4 M_{\odot}$ of mass
accreted via torus, which corresponds to the GRB duration 
of only about 8 seconds. 
In scenario {\it a}, the mass accreted with $l > l^{\rm k}_{\rm  crit}$ is
even smaller than that, especially for small $x$ ($ < 0.5$).
No mass will be accreted through the torus if $x\le 0.05$ (model {\bf C}) 
or $x\le 0.1$ (model {\bf D}).

In scenario {\it c}, the mass accreted with $l>l^{\rm k}_{\rm crit}$
is independent on $x$, if there is a cut off on $l_{\rm max}$ (cf. Eq.
\ref{eq:maccr} and note that this
mass is equivalent to the total mass accreted).

The accretion scenario {\it b}, i.e. that of the accretion composed from
two steps: polar funnel and than torus, is not discussed for models
{\bf C} and {\bf D}. It is because
only a very small fraction of the envelope, and for very small $x$,
has $l<l_{\rm crit}$, so basically the results will not differ
much from the scenario {\it c}.

 Finally, we tested the models {\bf C} and {\bf D} with an upper limit
imposed on the free fall timescale, $t_{\rm ff}^{\rm max}=1000$ s.
To comaper the two effects, in these tests 
we did not assume any limit for the specific angular momentum $l_{\rm max}$.
As shown in the upper panels in the Figures \ref{fig:figmcd} 
and \ref{fig:mcdtorus}, the mass accreted on the black hole
(both in total and via torus)
is now 3 times smaller than without the $t_{\rm ff}^{\rm max}$ limit.
Nevertheless, this effects is not as strong as the limit for the
progenitor rotation law imposed by $l_{\rm max}$.

To sum up, in the models {\bf C} and {\bf D}, i.e. if the specific
angular momentum in the pre-collapse star depends on $\theta$ and $r$
in such a way, that either the angular velocity $\Omega$ is constant,
or constant is the ratio between the gravitational and centrifugal
forces, a fraction of envelope material able to form a torus can be
much smaller, or much larger than in models {\bf A} and {\bf B}.
The fraction of 100\% is possible if there is no limiting value on
the specific angular momentum (or, more specifically,
 the limiting value exceeds 
$4.23 \times 10^{17}$ cm$^{2}$s$^{-1}$ in our model),
and no limit for a free fall timescale.
However, in more physical modeling which accounts for such limits,
 this fraction becomes very small:  
 for $t_{\rm ff}^{\rm max} = 1000$ s we obtain about 30\%
of the envelope accreted via torus, and 
for
$l_{\rm max} = 10^{17}$ cm$^{2}$s$^{-1}$ we obtain at most 15\%.

\subsection{Duration of a GRB}

\begin{figure}
\epsscale{.80}
\plotone{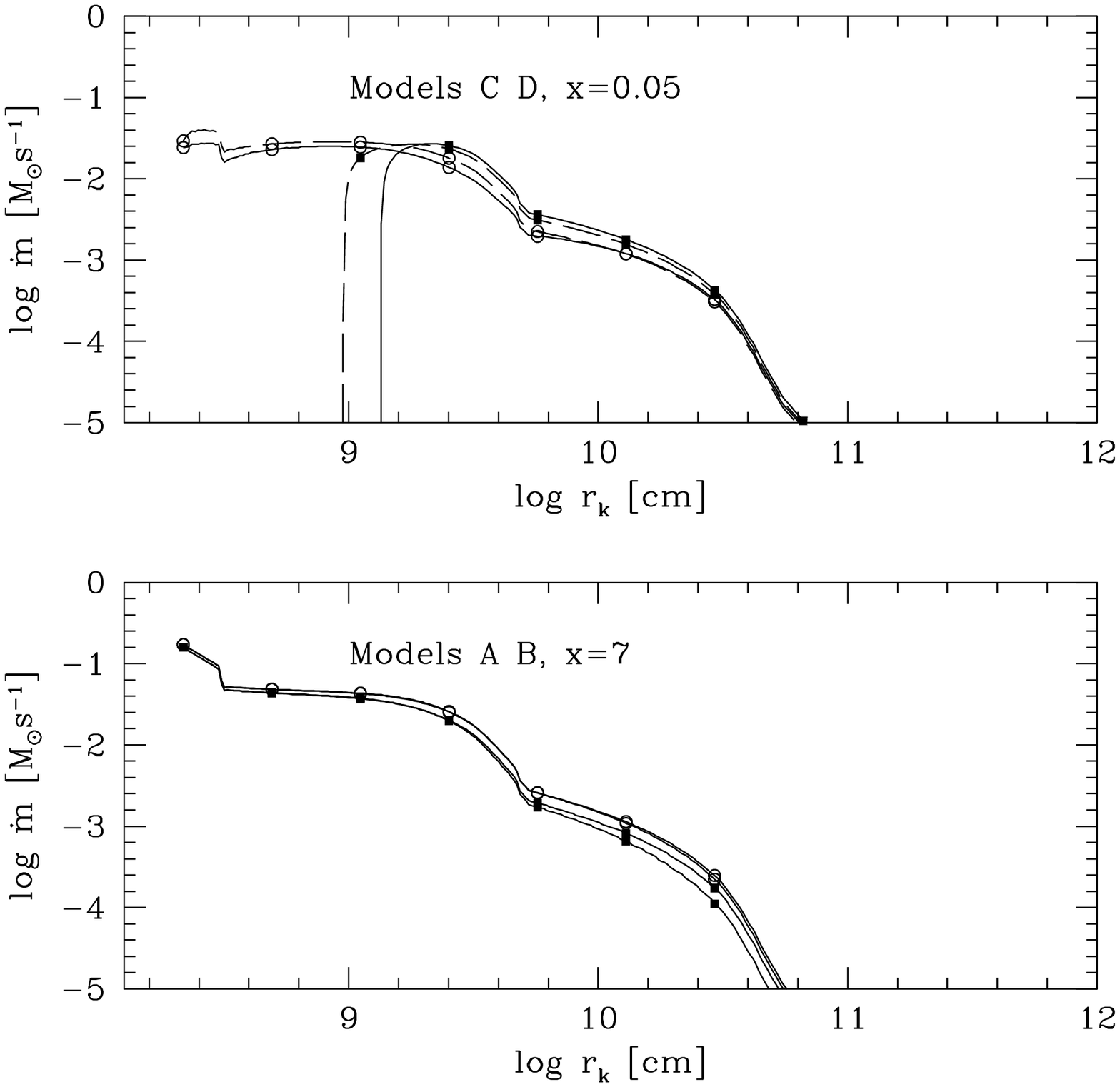}
\caption{
The instantaneous mass accretion rate during an iteration, for the 4 models
and 2 chosen values of the 
initial normalization of the specific angular momentum distribution.
The lower panel shows the models of the distribution function $f(r)$:
model {\bf A} (solid squares) and model {\bf B} (open circles) and
the upper panel shows the models of the distribution function $f(r)g(\theta)$: 
{\bf C} (solid squares) and {\bf D} (open circles).
The accretion scenarios are: {\it a} (solid line) and {\it c} (dashed line).
}
\label{fig:mdotinst}
\end{figure}

In Figure \ref{fig:mdotinst} we show the instantaneous accretion rate
during the iterations, i.e. as a function of the current radius.
As the Figure shows, the accretion rate is the largest at the beginning of 
the collapse, and equal about 0.1 $M_{\odot}$ s$^{-1}$.
In model {\bf C}, for $x=0.05$ the condition for torus formation 
is not satisfied initially (cf. Fig. \ref{fig:figlccd}), 
so the accretion rate through the torus is zero. For the same $x$, in model
{\bf D} the torus is already formed near the equatorial plane,
and the accretion rate is about 0.03 $M_{\odot}$ s$^{-1}$.

The duration of a gamma ray burst depends on the accretion rate and
the total mass accreted via torus onto the black hole.
If the torus is a poer source for a GRB, the accretion rate, 
albeit must not be constant, cannot drop to a very small value either.
Below, say 0.01 solar masses per second the neutrino cooling in the 
accretiong torus may become inefficient.
Table \ref{table:tgrb1} shows the results for the computations, 
in which we have limited the iterations to the minimum accretion rate of $\dot m_{\rm min} = 0.01 M_{\odot}$ s$^{-1}$. The table summarizes the mass acreted via torus for 
all of our models for the progeniotor rotation (i.e. {\bf A}, 
{\bf B}, {\bf C} and {\bf D}) as well as the two
accretion scenarios ({\it a} and {\it c}).
The mass accreted through the torus never exceeds 
4.5 $\dot M_{\odot}$ s$^{-1}$, thus implying that the limiting accretion rate
value influences the results stronger than the limits for the free-fall 
time, and comparably to the limit for progenitor rotation.

\begin{table}
\caption{
 The mass accreted through the torus (in solar masses)
for various models
and accretion scenarios, under the assumption that the minimum accretion rate
is $\dot m_{\rm min}=0.01 M_{\odot} s^{-1}$}
\begin{center}
\begin{tabular}{l c c c r }     
\hline\hline    
x & $M^{torus}_{Aa}$  & $M^{torus}_{Ac}$  & $M^{torus}_{Ba}$  & $M^{torus}_{Bc}$ \\
\hline
2.0 & 0.35 & 0.78 & 0.78 & 1.37 \\
3.0 & 1.01 & 1.58 & 1.76 & 2.37  \\
4.0 & 1.62 & 2.12 & 2.61 & 2.94  \\
5.0 & 2.14 & 2.54 & 3.09 & 3.27  \\
6.0 & 2.53 & 2.78 & 3.38 & 3.48  \\
7.0 & 2.82 & 3.02 & 3.57 & 3.64  \\
8.0 & 3.02 & 3.19 & 3.73 & 3.78  \\
9.0 & 3.19 & 3.30 & 3.81 & 3.85  \\
10.0& 3.33 & 3.43 & 3.88 & 3.91  \\
\hline
\hline 
x & $M^{torus}_{Ca}$ & $M^{torus}_{Cc}$ & $M^{torus}_{Da}$ & $M^{torus}_{Dc}$\\
\hline
0.05& 1.96 & 2.55 & 1.64 & 2.50 \\
0.1 & 2.93 & 3.18 & 3.41 & 3.53 \\
0.2 & 3.55 & 3.66 & 4.05 & 4.07 \\
0.3 & 3.80 & 3.85 & 4.21 & 4.23 \\
0.4 & 3.95 & 3.97 & 4.35 & 4.31 \\
0.5 & 4.04 & 4.09 & 4.39 & 4.35 \\
0.6 & 4.11 & 4.15 & 4.43 & 4.43 \\
0.7 & 4.17 & 4.21 & 4.45 & 4.45 \\
0.8 & 4.22 & 4.26 & 4.47 & 4.47 \\
0.9 & 4.27 & 4.30 & 4.48 & 4.48 \\
1.0 & 4.32 & 4.33 & 4.49 & 4.49 \\
\hline

\end{tabular}
\end{center}
\label{table:tgrb1}
\end{table}

The Table \ref{table:tgrb2} summarizes the durations of GRB prompt phase,
also for the four models and two accretion scenarios.
The results correspond to the total masses accreted via torus 
that are given in Table \ref{table:tgrb1}. The duration time
was calculated as $t_{\rm GRB} = M^{\rm torus}_{\rm acc} / <\dot m>$, 
where $<\dot m>$ is the mean accretion rate during an iteration.
Note, that because the minimum accretion rate was fixed at 
0.01 solar masses per second, the average value is equal to
$0.5 \dot m_{\rm max}$.

\begin{table}
\caption{Duration of GRB (in seconds) 
for various models
and accretion scenarios, under the assumption that the minimum accretion rate
is $\dot m_{\rm min}=0.01 M_{\odot} s^{-1}$}
\begin{center}
\begin{tabular}{l c c c r }     
\hline\hline    
x & $t_{Aa}$ & $t_{Ac}$ & $t_{Ba}$ & $t_{Bc}$  \\
\hline
2.0 &  7.00 & 15.23 & 11.17 & 19.75  \\
3.0 & 15.15 & 23.77 & 21.96 & 29.60  \\
4.0 & 21.96 & 28.86 & 30.72 & 34.68  \\
5.0 & 27.32 & 32.43 & 35.25 & 37.33  \\
6.0 & 31.01 & 34.08 & 37.80 & 38.93  \\
7.0 & 33.50 & 35.99 & 39.46 & 40.24  \\
8.0 & 35.28 & 37.34 & 40.91 & 41.49  \\
9.0 & 36.66 & 37.96 & 41.39 & 41.85  \\
10.0 & 37.70& 38.85 & 41.90 & 42.27  \\
\hline
\hline 
x & $t_{Ca}$ & $t_{Cc}$ & $t_{Da}$ & $t_{Dc}$\\
\hline
0.05& 106.25 & 139.75 & 87.84 & 99.89 \\
0.1 & 126.99 & 147.09 & 48.21 & 49.95 \\
0.2 & 136.21 & 150.19 & 47.62 & 47.70 \\
0.3 & 139.29 & 150.11 & 47.08 & 47.29 \\
0.4 & 141.32 & 150.17 & 47.41 & 46.99 \\
0.5 & 142.43 & 151.77 & 47.44 & 46.98 \\
0.6 & 143.26 & 151.66 & 47.32 & 47.37 \\
0.7 & 144.21 & 142.71 & 47.38 & 47.38 \\
0.8 & 131.66 & 126.07 & 47.22 & 47.20 \\
0.9 & 117.38 & 114.70 & 47.30 & 47.31 \\
1.0 & 108.81 & 106.52 & 47.26 & 47.28 \\
\hline

\end{tabular}
\end{center}
\label{table:tgrb2}
\end{table}

The calculated duration times of GRBs are the largest for models {\bf C}, 
because the average accretion rate in these models is the smallest.
Taking into account only the free fall timescale and 
under the adopted assumptions for a minimum $\dot m$, 
these models give at most $\sim 145$ s of the GRB prompt phase.
All the other models result in the GRB duration below 50 seconds.

\section{Discussion and conclusions}
\label{sec:diss}

The durations of  GRBs range from less than 0.01 to a few hundred seconds 
(for a review see e.g. Piran 2005), and the long duration
bursts,
$T_{90}> 2$ s, are supposed to originate from the collapse of the
massive rotating star. The collapsar model assumes 
that the presence of a rotating
torus around a newly born black hole is a crucial element of the GRB
central engine for the whole time of its duration. 
In this work, we found that 
 some specific properties of the progenitor star are important in
 order to support the existence of a torus, which consists of the
material with specific angular momentum larger than a critical value 
$l>l_{\rm crit}$.
We studied how the initial distribution of specific angular momentum
inside the stellar envelope affects the burst duration, taking into 
account the increase of the black hole mass 
during the collapse process.

Following Woosley \& Weaver (1995),
we considered  the model of pre-supernova star that predicts the existence of
the $\sim 1.7 M_{\odot}$ iron core, which forms a black hole, and the
$\sim 24 M_{\odot}$ envelope. Therefore in the simplest approach, when
the mass available for accretion is the total envelope mass, and
the accretion rate is a constant value on the order of 
$0.1-0.5 M_{\odot}$s$^{-1}$, the
central engine is able to operate for a time required to power a long
GRB (i.e. several tens to a couple of hundreds of seconds).

However, McFadyen \& Woosley (1999) in their collapsar model show that
most of the initial accretion goes through
the rotating torus rather than from the polar funnel. Torus formation 
is possible if
material with substantial specific angular momentum is present in the envelope,
initially and throughout  the event as well. In this
sense the GRB duration estimated in the uniform  accretion scenario
is only an upper limit.

In our calculations, this upper limit is achieved only 
if the specific angular momentum distribution in the 
pre-supernova star is a strong function of radius (i.e. 
$g(r)\sim r^{2}$ or $g(r)\sim \sqrt{r}$), and the inner parts have
$x = l/l_{\rm crit}(r_{\rm in}) \sim 1.0$ 
(for the initial black hole mass we have 
$l_{\rm crit}^{0} \sim 3\times 10^{16}$ cm$^{2}$s$^{-1}$ in our model),
while the 
outer parts of the star may have an specific angular momentum as large as
$l_{\rm max} \ge 4.23\times 10^{17}$ cm$^{2}$s$^{-1}$.
Both these conditions challenge the progenitor star models:
they require either a rigid rotation of the star, or a huge ratio of
centrifugal
to gravitational force. The latter, if we want to keep the value
of $F_{\rm centr}/F_{\rm graw} \sim 0.02$ (as taken by
Proga et al. 2003), would lead to the mass accreted through the torus
equal to be a fraction of the envelope mass, and a
correspondingly shorter GRB duration (i.e. one should take $x=0.05$,
which implies the
the GRB duration in our 'mass units' to be 20-21 $M_{\odot}$, cf. 
Fig. \ref{fig:mcdtorus}).

Furthermore, the progenitor star models 
used in MacFadyen \& Woosley (1999; see also Proga et al. 2003), would rather
assume a limiting value for the specific angular momentum.
In our modeling we followed that work and calculated an exemplary
sequence of models with $l_{\rm max} = 10^{17}$
cm$^{2}$s$^{-1}$. 
The results in this case are not promising for a long GRB production:
the GRB duration in the accreted mass units does not exceed
4 $M_{\odot}$, which would give an event not longer than a hundred
 seconds and only if the accretion rate is very small.

The models with  the specific angular momentum distribution depending only
on the polar angle, $\theta$, also yield not very long GRB
durations. In this case, the mass accreted with $l>l_{\rm crit}$ is of 
the order of 15 $M_{\odot}$ if the specific angular momentum in 
the progenitor star is about ten times the critical one (i.e. $x \sim 10$), 
and the accretion proceeds only
through the torus (the latter finds support in MHD simulations
such those performed by Proga et al. 2003).
If the accretion proceeds uniformly through the envelope, 
the GRB duration drops to about 10 $M_{\odot}$ for the same value of $x$.
Finally, in the scenario when accretion proceeds first through the
poles and then through the torus (i.e. scenario {\it b}
as indicated by HD simulations performed by MacFadyen \& Woosley 1999),
there is no GRB for $x\le 5-7$ (depending on the shape of
$f(\theta)$), because all the material is swollen by the black hole
during the first stage. For large $x$, the resulting GRB duration
is in between of the scenarios {\it a} and {\it c}.
We plan to consider other progenitor star models such as those
computed by Heger et al. (2005)  to check how our conclusions
depend on specific radial and rotational profiles.

We also investigated the models in which the mass accreted onto BH
was limited by the free fall timescale or the minimum accretion rate.
In case of of the free fall time limited to 1000 seconds,
the mass accreted onto the black hole is much smaller than the total 
envelope mass, and reached up to 8 $M_{\odot}$ but for a very 
fast rotation of the progenitor star. 
Finally, the explicitly calculated duration times of GRB, 
obtained due to the released assumption of a constant accretion rate,
were at most 30-150 seconds, depending on the model of the specific 
angular momentum distribution and accretion scenario.

The effects of accreting non-rotating or slowly rotating
matter on a black hole can reduce also capability of
powering GRBs through the Blandford-Znajek mechanism.
In the estimations done by McFadyen \& Woosley (1999) a-posteriori,
i.e. using the analytical models of the accretion disk 
to extend the calculations down to the
event horizon, the authors calculated
the evolution of the black hole mass and angular momentum.
The initial dimensionless angular momentum parameter 
of the iron core is taken in their work to be either
$a_{\rm init}=0$ or $a_{\rm init}=0.5$.
However, the black hole changes its total mass
and angular momentum as it accretes matter (see Bardeen 1977 for
specific formulae). 
In this way, if the specific angular momentum supply 
is substantial, even starting from $a=0$, a Schwarzschild black hole, 
the finite amount of accreted mass makes it possible to obtain $a=1$.
On the other hand, the material
with very small specific angular momentum, which is present
in a collapsing star, will spin down the black hole.
 
The effect of the evolution of the black hole spin due to the
accretion (spin up) and the Blandford-Znajek mechanism (spin down)
has been studied in Moderski \& Sikora (1996).
Lee, Brown \& Wijers (2002) studied the case of GRBs produced after the progenitor star
has been spun up in a close binary system due to spiral-in and tidal coupling.
Recently, Volonteri et al. (2005) and
King \& Pringle (2006) calculated the spin evolution of supermassive
black holes due to the random accretion episodes.
In our model, the black hole spin evolution
is not episodic, but a continuous process. The calculations of the
BH angular momentum evolution are left for the future work.
Such calculations may possibly show that obtaining a large BH spin
parameter, $a\sim 0.9$, is rather difficult when a large fraction of
the envelope material has $l \ll l_{\rm crit}$.

\section*{Acknowledgments}
We thank Phil Armitage for useful comments.
This work was supported by NASA under grants NNG05GB68G.

\end{document}